\documentclass{PoS}

\graphicspath{{.}{figs/}}
\usepackage{epstopdf}
\DeclareGraphicsExtensions{.png, .jpg, .pdf, .eps}
\usepackage{aas_macros}

\title{DASCH 100-yr light curves of high-mass X-ray binaries}

\ShortTitle{DASCH 100-yr light curves of high-mass X-ray binaries}

\author{\speaker{Mathieu Servillat}
         \thanks{ The DASCH project is supported in part by the NSF grants AST0407380 and AST0909073 and the Cornel and Cynthia K. Sarosdy Fund for DASCH. MS acknowledges founding from the Centre National d'Etudes Spatiales (CNES).}\\
        Laboratoire AIM (CEA/DSM/IRFU/SAp, CNRS, Universit\'e Paris Diderot), CEA Saclay, Bat.~709, 91191 Gif-sur-Yvette, France\\
        E-mail: \email{mathieu.servillat@cea.fr}}

\author{S. Tang, J.E. Grindlay, E. Los\\
        Harvard College Observatory, 60 Garden Street, Cambridge, MA 02138, USA\\}

\abstract{
We analyzed the 100-yr light curves of Galactic high-mass X-ray binaries using the Harvard photographic plate collection, made accessible through the DASCH project (Digital Access to a Sky Century at Harvard). As scanning is still in progress, we focus on the four objects that are currently well covered: the supergiant X-ray binary Cyg X-1 (V1357 Cyg), and the Be X-ray binaries 1H~1936+541 (BD+53 2262), RX~J1744.7-2713 (HD 161103), and RX J2030.5+4751 (SAO 49725), the latter two objects being similar to $\gamma$ Cas. 
The star associated with \mbox{Cyg X-1} does not show evidence for variability with an amplitude higher than 0.3 magnitude over a hundred years.
We found significant variability of one magnitude with timescales of more than 10~years for SAO~49725, as well as a possible period of $\sim$500--600 days and an amplitude of 0.05 magnitude that might be the orbital, or super-orbital period of the system. The data is insufficient to conclude for HD 161103 but suggests a similar long-term variability. We thus observe an additional characteristic of $\gamma$ Cas-like objects: their long-term variability. 
This variability seems to be due to the slow evolution of a decretion disk around the Be star, but may be triggered by the presence of a compact object in the system, possibly a white dwarf.
This characteristic could be used to identify further similar objects otherwise difficult to detect. 
}

\FullConference{ An INTEGRAL view of the high-energy sky (the first 10 years) - 9th INTEGRAL Workshop and celebration of the 10th anniversary of the launch\\
                 15-19 October 2012\\
                 Bibliotheque Nationale de France, Paris, France}

\begin{document}

\section{Introduction}

High-mass X-ray binaries (HMXBs) are composed of a compact object (a neutron star or a black hole) that accretes material from a massive early-type O-B star. Conventionally, they are subdivided into two groups, the supergiant X-ray binaries (sgXBs) and Be/X-ray binaries (BeXBs). 

In sgXBs, the massive star (type I/II star) ejects a slow, dense wind and the compact object directly accretes the stellar wind, leading to a persistent X-ray emission ($L_{X}\sim10^{35-36}$~erg~s$^{-1}$). They exhibit rare Type II outbursts and no Type I outbursts. 
Some sgXBs are Roche lobe overflow systems causing a high X-ray luminosity ($L_{X}\sim10^{38}$~erg~s$^{-1}$) during outbursts. In particular, \mbox{Cyg X-1} is the only sgXB with Roche lobe overflow and stellar wind accretion hosting a confirmed black hole. It is generally found in the low state.

A BeXB consists of a neutron star orbiting a Be star in a wide (period of months) and eccentric orbit. 
The Be star is surrounded by a disc of relatively cool material, presumably ejected from the star due to causes unknown, but generally believed to be associated with fast rotation, magnetic fields and/or non-radial pulsations \cite{Slettebak:1988p9009}.
In BeXB systems, the neutron star can penetrate this equatorial disc during each periastron passage, giving rise to periodic outbursts (Type I) that can be observed over a wide range of wavelengths (optical, X-ray, ...).
Be stars are variable and may lose their disk (and hence emission) completely and be classified as normal B-type stars. Therefore, census studies will provide a lower limit to the fraction of Be stars within a sample.
A scenario for B stars to become Be stars proposes the rapid rotation of Be stars as consequence of binary evolution spin-up \cite{Pols:1991p10371}. Hence all Be stars would have evolved companions. 
The Be star is hardly affected in a BeXB: while the disks' radii have an effective upper size, probably determined by the binary parameters \cite{Zamanov:2001p10396}, they are otherwise not significantly different from single Be star disks.
A recent review of basic properties of BeXB can be found in \cite{Reig:2011p9029}.

A growing number of early Be stars discovered in X-ray surveys with low-luminosity hard X-ray emission ($L_{X}\sim10^{32-33}$~erg~s$^{-1}$) show a very likely thermal emission of high temperature ($kT\sim8$~keV). 
A strong iron line, including a fluorescence component, was detected for some systems. 
They also share very similar optical properties such as spectral types (B0.5III-Ve) and strength of the Balmer emission. 
This pattern of common X-ray and optical properties appears quite similar to that of $\gamma$~Cas and points to the emergence of a new class of $\gamma$~Cas analogs \cite{LopesdeOliveira:2006p7786}.
Two models have been put forward to explain this type of system: accretion onto a compact object (most likely a white dwarf) and magnetically heated material between the photosphere of the B star and the inner part of its disc \cite{Robinson:2002p9097}.



The Digital Access to a Sky Century at Harvard (DASCH) is a project to digitize and analyze the scientific data contained in the $\sim$530\,000 Harvard College Observatory (HCO) plates taken between the 1880s and 1990s, which is a unique resource for studying temporal variations in the universe on 10--100 yr timescales \cite{Grindlay:2009p626}. We have developed the astrometry and photometry pipeline \cite{Laycock:2010p742,Los:2011p8977,Servillat:2011p10415}. We started to scan plates in several selected fields and discovered new types of variable stars \cite{Tang:2010p3252,Tang:2011p7922,Tang:2012p7113}.

We initiate here the study of the long-term variability of known HMXBs in the Galaxy by studying the first four objects covered by the current DASCH scanned images.
DASCH will enable a large study of the long-term variability of Galactic HMXBs and the brightest HMXBs in the Magellanic Clouds. 

\section{Data}
\label{data}

\subsection{Search for HMXBs in DASCH data}

\begin{table}
\small
\centering
\begin{tabular}{llcccc}
\hline
{X-ray source name} & {Star name} & {Spectral type} & {Bmag} & {RA} & {Dec} \\
\hline
RX J1744.7-2713 & HD 161103  & B0.5III-Ve  & 9.13 & 17:44:45.70 & -27:13:44.0 \\
1H 1936+541       & BD+53 2262 & Be        & 10.38 & 19:32:52.30 & +53:52:45.0 \\
Cyg X-1 / 4U 1956+3 & V1357 Cyg & O9.7Iab & 9.62 & 19:58:21.70 & +35:12:06.0 \\
RX J2030.5+4751 & SAO 49725 & B0.5III-Ve & 9.57 & 20:30:30.80 & +47:51:51.0 \\
\hline
\end{tabular}
\caption{HMXB candidates with DASCH data.}
\label{tab:hmxbs}
\end{table}

The DASCH database now contains measurements from 38\,000 plates (November 2012), 7\% of the total collection, covering mainly five selected fields in the sky.
As an initial study of the HMXB population, we used the catalog of HMXBs from \cite{Liu:2006p3344} and searched DASCH for matches. 
This catalog contains 114 HMXBs, however, some sources are only tentatively identified as HMXBs on the basis of their X-ray properties similar to the known HMXBs.
About 60\% of the catalogued HMXB candidates are known or suspected BeXB, while 32\% are sgXB.  

We found that four objects have DASCH light curves with more than 500 magnitude measurements over 100 years. They are listed in Table~\ref{tab:hmxbs} with their spectral type, coordinates and current B magnitude (Bmag) as found in SIMBAD\footnote{The SIMBAD database is operated at CDS, Strasbourg, France (http://simbad.u-strasbg.fr/simbad)}. One of those objects is the sgXB Cyg X-1, and the three others are BeXB. We note that two objects (HD 161103 and SAO 49725) are peculiar HMXB candidates similar to $\gamma$ Cas for which the binary nature remains to be proven \cite{LopesdeOliveira:2006p7786}.

\subsection{Cleaning of the light curves}

\begin{figure}
\centering
\includegraphics[width=.45\columnwidth]{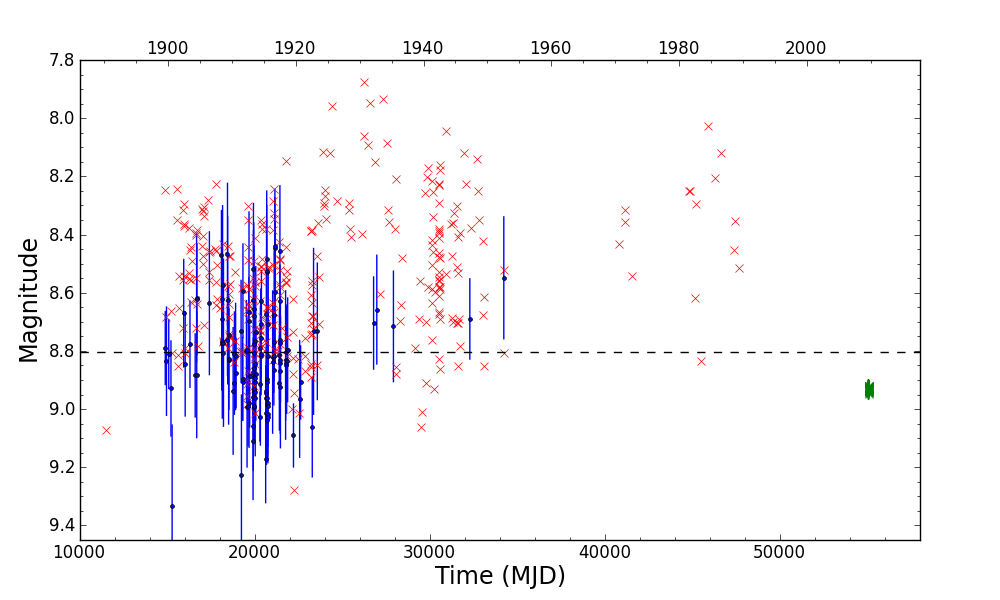} 
\includegraphics[width=.45\columnwidth]{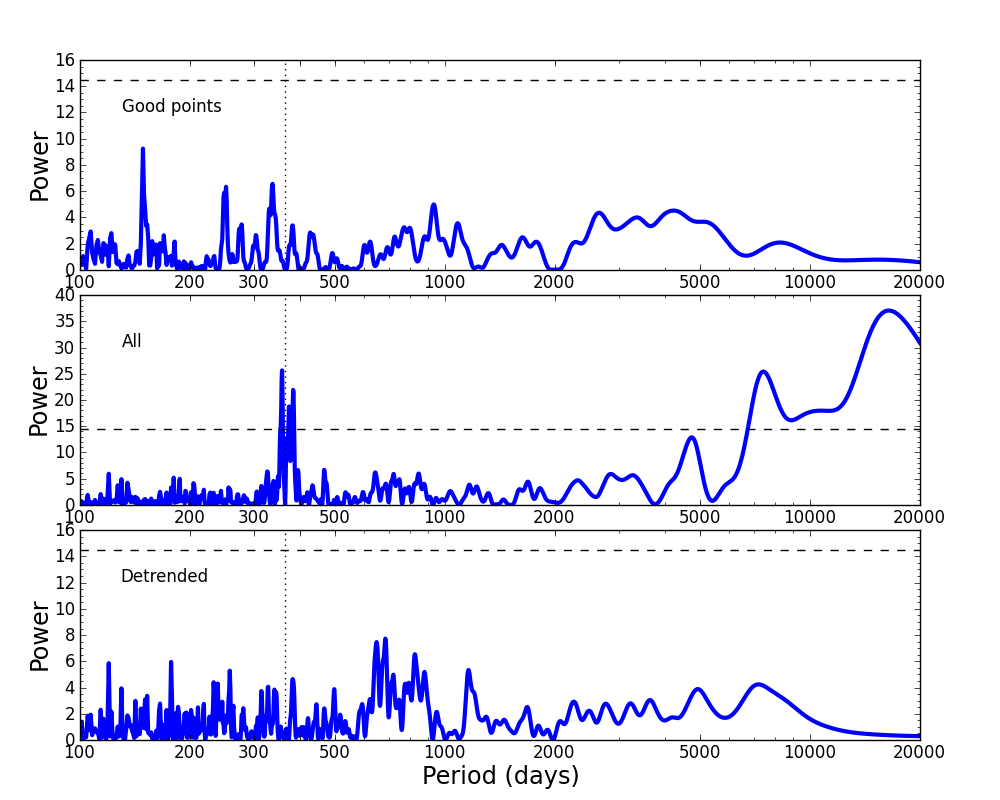} 
\caption{HD 161103 light curve and periodograms for DASCH good points, all points (including saturated points shown as red crosses in the light curve) and for the detrended light curve.\label{fig:lc1}}
\end{figure}

\begin{figure}
\centering
\includegraphics[width=.45\columnwidth]{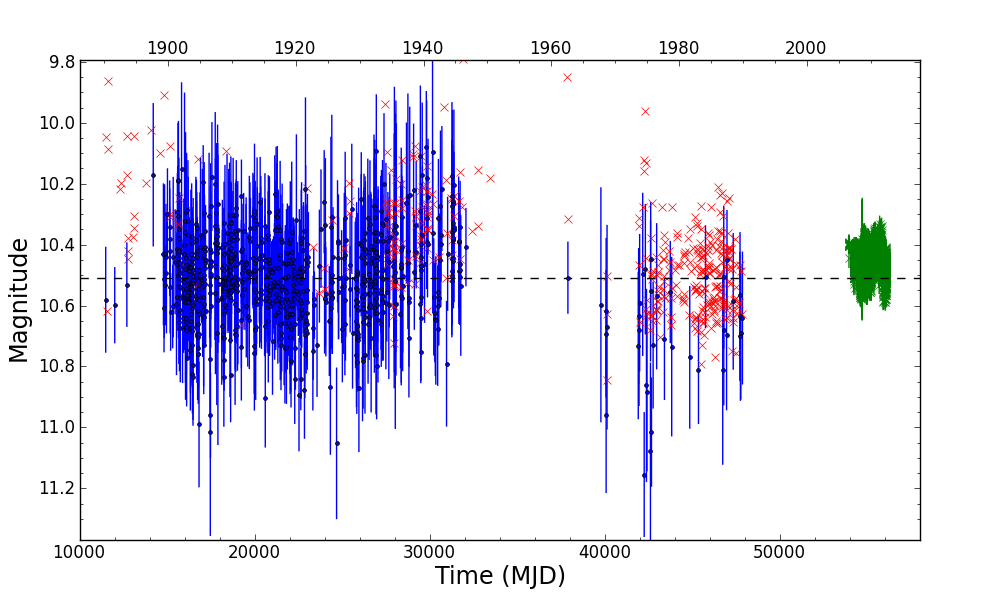}
\includegraphics[width=.45\columnwidth]{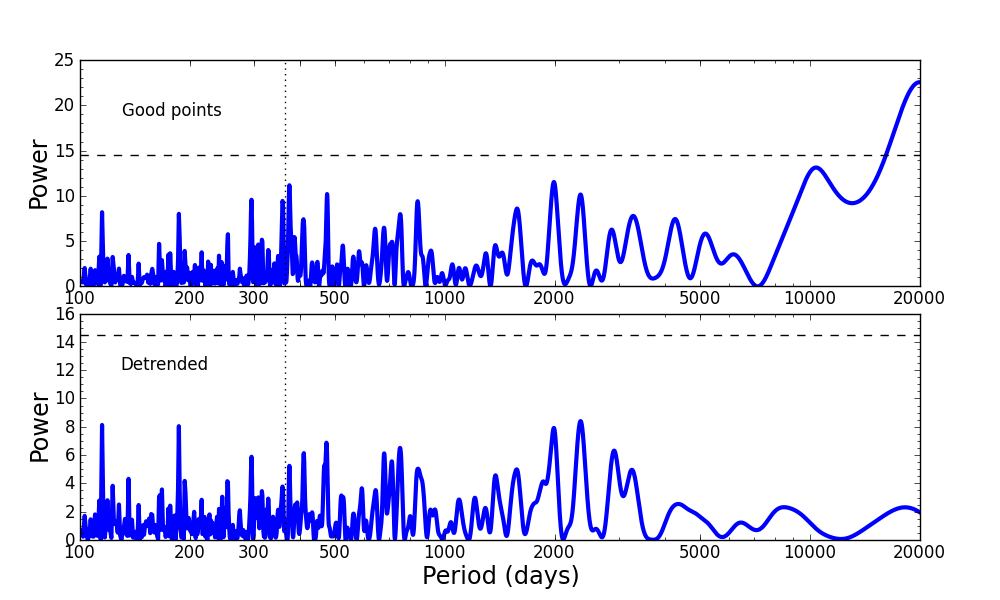} 
\caption{BD+53 2262 light curve and periodograms. The AAVSO points are shown in green.\label{fig:lc2}}
\end{figure}

\begin{figure}
\centering
\includegraphics[width=.45\columnwidth]{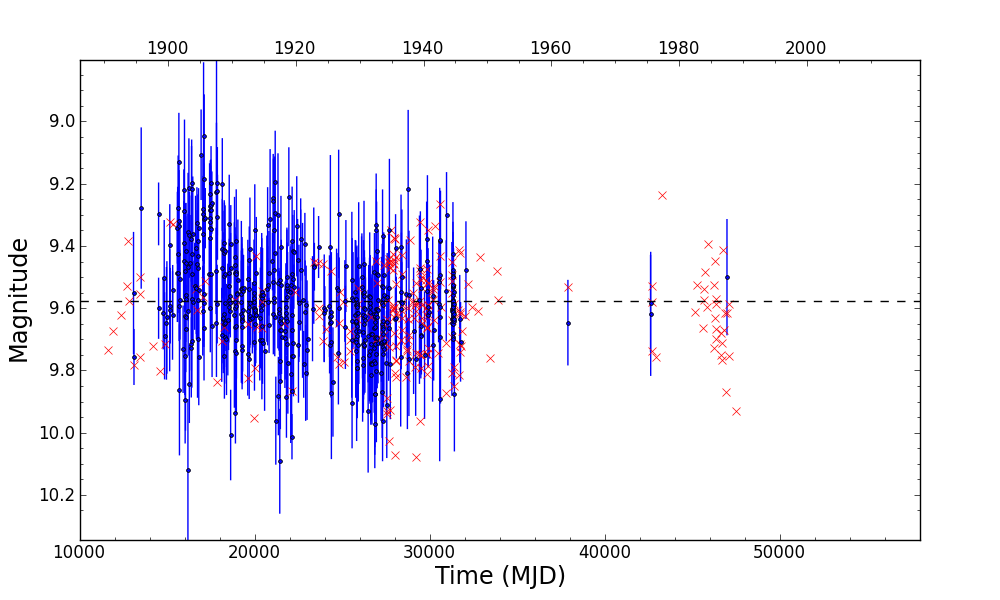}
\includegraphics[width=.45\columnwidth]{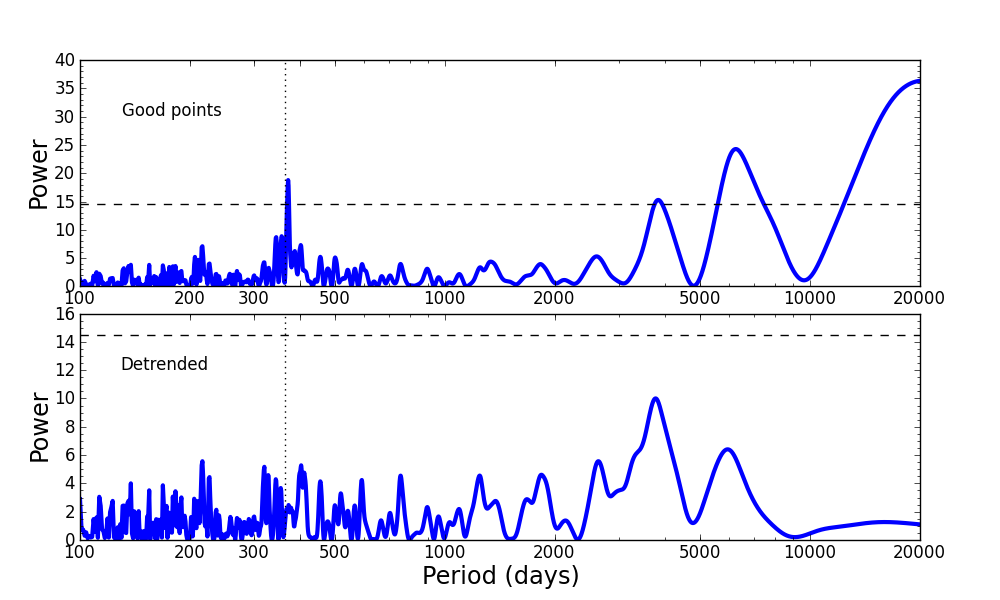} 
\caption{V1357 Cyg light curve and periodograms.\label{fig:lc3}}
\end{figure}

\begin{figure}
\centering
\includegraphics[width=.45\columnwidth]{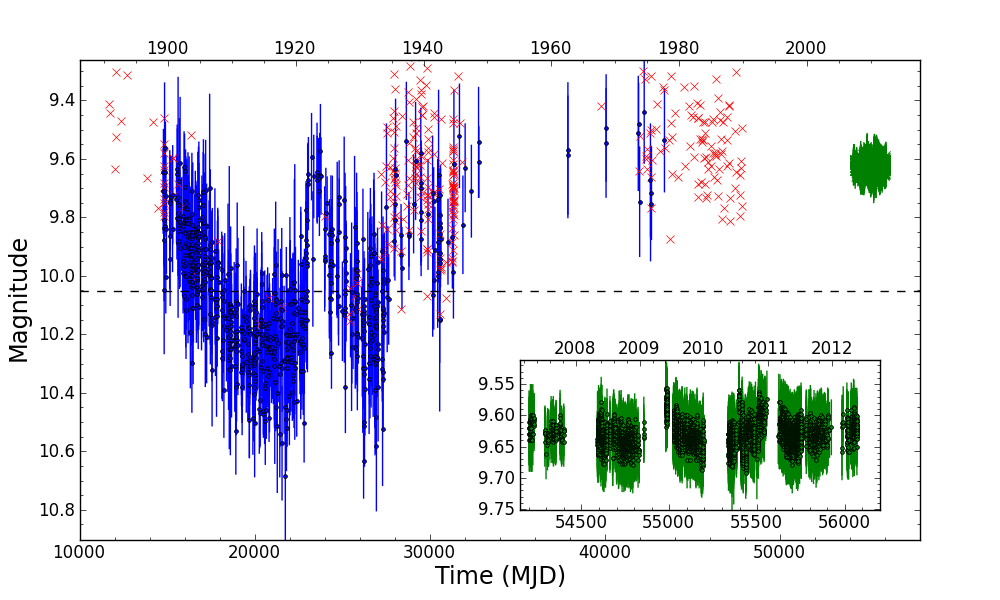} 
\includegraphics[width=.45\columnwidth]{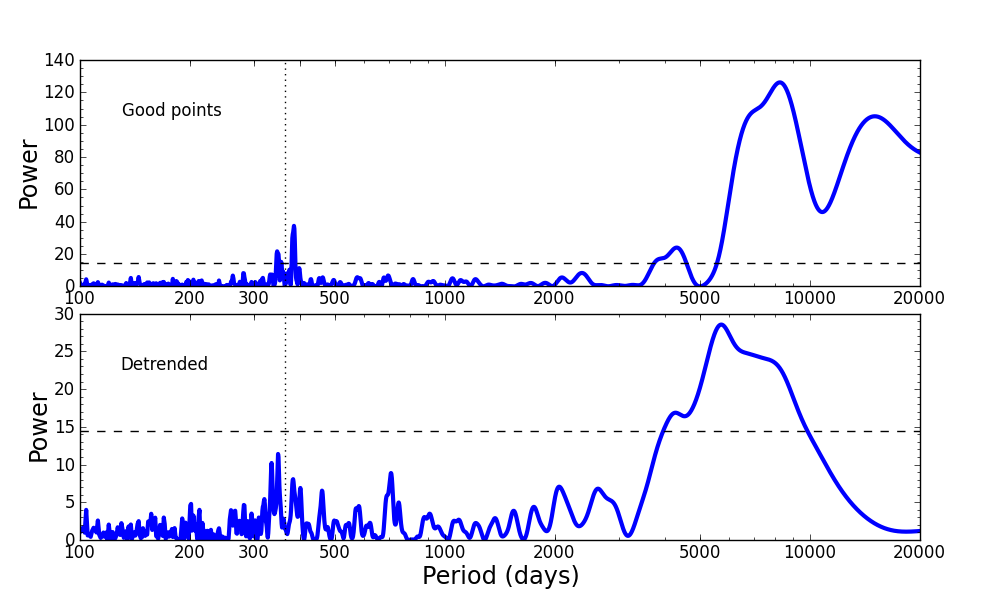} 
\caption{SAO 49725 light curve and periodograms. The AAVSO points are shown in green.\label{fig:lc4}}
\end{figure}

We removed all points from plates flagged as showing defects, and we performed a 3$\sigma$ clipping to remove isolated outliers in order to focus on long-term variability. Points flagged as saturated are plotted with crosses and might be affected by calibration issues, though they may contain relevant information. The four light curves listed in Table~\ref{tab:hmxbs} are shown in Figures~\ref{fig:lc1},~\ref{fig:lc2},~\ref{fig:lc3}, and~\ref{fig:lc4} (left panel).

Cyg X-1 = V1357 Cyg is close to V1674 Cyg (55''). Both are variable stars with B mean magnitudes 9.62 and 10.6, respectively. For some plate series with low resolution (6'' per pixel), the two stars can be blended. 
The brightest magnitudes observed in the uncleaned light-curve (9.0) matches well the expected magnitude of the two stars blended together, raising doubts on any real variability detected in the DASCH light curve.

To complement the DASCH light curves with data after 2000, we downloaded the available data from the AAVSO\footnote{American Association of Variable Star Observers (http://www.aavso.org)} website for those objects. We used the AAVSO B magnitudes only, given that DASCH magnitude measurements are similar to B magnitudes. 

\subsection{Periodicity search}



We only kept the non-saturated points and computed a Lomb-Scargle periodogram following the method described in \cite{Townsend:2010p9111}. We first subtracted the mean flux in the light curve and set the variance to one to obtain the normalized periodogram. We searched for periods between 100 and 20\,000 days ($\sim55$~years) given the typical sampling and duration of DASCH light curves. The light curves and periodograms are shown in Figures~\ref{fig:lc1},~\ref{fig:lc2},~\ref{fig:lc3}, and~\ref{fig:lc4} (left and right panels, respectively).

We used a local regression technique (lowess) with a window of 7\,000 days (about 20 years) to extract the long-term variability from the light curves. We obtained a detrended light curve after dividing the initial light curve by this smoothed light curve. We then recomputed a periodogram for the detrended light curve. 

We note that there is often a signal around 365 days (note also the weak signal at 180 days for HD 161103, Figure~\ref{fig:lc1}). This one year period seems to be due to the long-term variability of an unevenly sampled light curve. This phenomenon is due to spectral leakage of low-frequency power present in the light curves (see e.g. \cite{Farrell:2005p9408}).
Indeed, the high power at one year disappears when the light curve is detrended, while a real signal would be strengthened. 

There is no significant variability in the light curves of HD 161103, BD+53 2262 and V1357 Cyg. No feature is observed in the detrended periodograms of those stars, indicating that no periodicity is observed. Cyg X-1 has an orbital period of 5.6 days and showed ellipsoidal variations with a low amplitude of $\sim$0.05 magnitude \cite{Walker:1972p10612}. Here we can rule out modulations with an amplitude higher than $\sim$0.3 magnitude. For HD 161103, we note that many points are flagged as saturated, but the light curve might also indicate large time scale variability.

The variability of SAO 49725, almost one magnitude on a timescale of more than 10 years, is clearly seen in the light curve (Figure~\ref{fig:lc4}, left). We find a signal around 5\,000 days ($\sim$15 years) in the periodogram which can be seen in Figure~\ref{fig:lc4} (two cycles from 1915 to 1945). However, the periodicity is not seen in the first part of the light curve (1900 to 1915), indicating that this variability is not periodic. 
For the AAVSO light curve of SAO 49725, we find a possible signal around 500--600 days. Indeed, two cycles can be seen in Figure~\ref{fig:lc4} (inset in left panel), with peaks around MJD 55\,000 and 55\,500, and an amplitude of $\sim$0.05 magnitude. This kind of period is difficult to detect in the DASCH light curve due to the low amplitude and larger error bars.


%
%

\section{Discussion and conclusions}
\label{discuss}

For the first time, we study a sample of HMXBs over a hundred years using optical light curves from the DASCH project. This study reveals a new kind of variability, poorly studied for HMXBs. A more complete study will follow when the Galactic plane will be fully covered by DASCH.

We do not expect to detect large variability amplitudes from sgXB, as illustrated by the case of Cyg X-1, because the optical light is mostly emitted by the supergiant star. However, Be stars can be extremely variable over a hundred years, as shown in Figure~\ref{fig:lc4} for SAO 49725.


It is interesting to note that among the stars presented in this study, the two sources that are found to be highly variable (clear variability for SAO 49725, to be confirmed for HD 161103) are $\gamma$~Cas-like objects.
This may be a remarkable property of those objects, as it reminds the long-term variability of $\gamma$~Cas itself, that varied from 1.6 to 3.0 magnitudes in the V-band between 1935 and 1940, and then slowly brightened to 2.1 today.
The long-term, non-periodic variability we observe for SAO~79425 could be due to changes in the decretion disk properties (size, density, temperature).
We found a weak hint of a period around 500-600 days, consistent with an orbital or a super-orbital period of a BeXB (e.g. \cite{Rajoelimanana:2011p6008}). This would indicate a possible link between binarity and long-term variability. 
A similar kind of variability has been observed for some neutron stars BeXBs in the Magellanic Clouds \cite{Rajoelimanana:2011p6008}.
However, a clear proof of binarity is still missing for $\gamma$~Cas-like objects.
It is still not clear if they are isolated Be stars, or if they host a compact object, probably a white dwarf to explain the hard X-ray spectrum with intermediate luminosities between Be stars and BeXB \cite{LopesdeOliveira:2006p7786}.


For the next few years, with a complete coverage of the Galactic plane, DASCH brings the prospect of detection of new HMXBs through the detection of long-term variability of B/Be~stars.



\bibliographystyle{JHEP} 
\bibliography{../../papers/ref.bib}

\providecommand{\href}[2]{#2}\begingroup\raggedright\begin{thebibliography}{10}

\bibitem{Slettebak:1988p9009}
A.~Slettebak, {\it The be stars},  {\em \pasp} {\bf 100} (Jul, 1988) 770.

\bibitem{Pols:1991p10371}
O.~R. Pols, J.~Cote, L.~B. F.~M. Waters, and J.~Heise, {\it The formation of be
  stars through close binary evolution},  {\em \aap} {\bf 241} (Jan, 1991) 419.

\bibitem{Zamanov:2001p10396}
R.~K. Zamanov, P.~Reig, J.~Mart{\'\i}, M.~J. Coe, J.~Fabregat, N.~A. Tomov, and
  T.~Valchev, {\it Comparison of the H-alpha circumstellar disks in be/x-ray
  binaries and be stars},  {\em \aap} {\bf 367} (Mar, 2001) 884.

\bibitem{Reig:2011p9029}
P.~Reig, {\it Be/x-ray binaries},  {\em Astrophys. Space Sci.} {\bf 332} (Mar,
  2011) 1.

\bibitem{LopesdeOliveira:2006p7786}
R.~{Lopes~de~Oliveira}, C.~Motch, F.~Haberl, I.~Negueruela, and
  E.~Janot-Pacheco, {\it New gamma Cas-like objects: X-ray and optical
  observations of SAO 49725 and HD 161103},  {\em \aap} {\bf 454} (Jul, 2006)
  265.

\bibitem{Robinson:2002p9097}
R.~D. Robinson, M.~A. Smith, and G.~W. Henry, {\it X-ray and optical variations
  in the classical be star γ cassiopeia: The discovery of a possible magnetic
  dynamo},  {\em \apj} {\bf 575} (Aug, 2002) 435.

\bibitem{Grindlay:2009p626}
J.~Grindlay, S.~Tang, R.~Simcoe, S.~Laycock, E.~Los, D.~Mink, A.~Doane, and
  G.~Champine, {\it Dasch to measure (and preserve) the harvard plates: Opening
  the 100-year time domain astronomy window},  {\em In: Preserving Astronomy's
  Photographic Legacy. ASP Conf. Series} {\bf 410} (Aug, 2009) 101.

\bibitem{Laycock:2010p742}
S.~Laycock, S.~Tang, J.~Grindlay, E.~Los, R.~Simcoe, and D.~Mink, {\it Digital
  access to a sky century at Harvard: Initial photometry and astrometry},  {\em
  \aj} {\bf 140} (Oct, 2010) 1062.

\bibitem{Los:2011p8977}
E.~Los, J.~Grindlay, S.~Tang, M.~Servillat, and S.~Laycock, {\it The DASCH data
  processing pipeline and multiple exposure plate processing},  {\em ADASS XX,
  ASP Conf. Ser.} {\bf 442} (Jul, 2011) 269.

\bibitem{Servillat:2011p10415}
M.~Servillat, E.~J. Los, J.~E. Grindlay, S.~Tang, and S.~Laycock, {\it
  Correcting the astrometry of DASCH scanned plates},  {\em ADASS XX, ASP Conf.
  Ser.} {\bf 442} (Jul, 2011) 273.

\bibitem{Tang:2010p3252}
S.~Tang, J.~Grindlay, E.~Los, and S.~Laycock, {\it DASCH discovery of large
  amplitude 10-100 year variability in K giants},  {\em \apjl} {\bf 710} (Feb,
  2010) L77.

\bibitem{Tang:2011p7922}
S.~Tang, J.~Grindlay, E.~Los, and M.~Servillat, {\it DASCH on ku cyg: A 5
  year dust accretion event in ~ 1900},  {\em \apj} {\bf 738} (Sep, 2011) 7.

\bibitem{Tang:2012p7113}
S.~Tang, J.~E. Grindlay, M.~Moe, J.~A. Orosz, R.~L. Kurucz, S.~N. Quinn, and
  M.~Servillat, {\it DASCH discovery of a possible nova-like outburst in a
  peculiar symbiotic binary},  {\em \apj} {\bf 751} (Jun, 2012) 99.

\bibitem{Liu:2006p3344}
Q.~Z. Liu, J.~van Paradijs, and E.~P.~J. van~den Heuvel, {\it Catalogue of
  high-mass X-ray binaries in the galaxy (4th edition)},  {\em \aap} {\bf 455}
  (Sep, 2006) 1165.

\bibitem{Townsend:2010p9111}
R.~H.~D. Townsend, {\it Fast calculation of the lomb-scargle periodogram using
  graphics processing units},  {\em \apjs} {\bf 191} (Dec, 2010) 247.

\bibitem{Farrell:2005p9408}
S.~A. Farrell, P.~M. O'Neill, and R.~K. Sood, {\it Recurrent 24h periods in
  RXTE ASM data},  {\em PASA} {\bf 22} (Aug, 2005) 267.

\bibitem{Walker:1972p10612}
E.~N. Walker, {\it B and v photometry of Cygnus X-1},  {\em \mnras} {\bf 160}
  (Jan, 1972) 9P.

\bibitem{Rajoelimanana:2011p6008}
A.~Rajoelimanana, P.~A. Charles, and A.~Udalski, {\it Very long-term optical
  variability of high-mass X-ray binaries in the small magellanic cloud},  {\em
  \mnras} {\bf 413} (May, 2011) 1600.

\end{thebibliography}\endgroup

\end{document}